\documentclass[aps,prc,
superscriptaddress,
preprint,
showpacs]{revtex4-1}
\usepackage{amsmath}
\usepackage{color}
\usepackage{graphicx}
\begin{document}
\title{Investigation of ${}^{9}$Be from nonlocalized clustering concept}
\author{Mengjiao Lyu} \email{mengjiao\_lyu@hotmail.com.}
\affiliation{School of Physics and Key Laboratory of Modern Acoustics,
Institute of Acoustics, Nanjing University, Nanjing 210093, China}
\affiliation{Research Center for Nuclear Physics (RCNP), Osaka
  University, Osaka 567-0047, Japan}

\author{Zhongzhou Ren} \email{zren@nju.edu.cn.}  \affiliation{School
  of Physics and Key Laboratory of Modern Acoustics, Institute of
  Acoustics, Nanjing University, Nanjing 210093, China}
\affiliation{Center of Theoretical Nuclear Physics, National
  Laboratory of Heavy-Ion Accelerator, Lanzhou 730000, China}

\author{Bo Zhou} \email{bo@nucl.sci.hokudai.ac.jp} \affiliation{School
  of Physics and Key Laboratory of Modern Acoustics, Institute of
  Acoustics, Nanjing University, Nanjing 210093, China}
\affiliation{Meme Media Laboratory, Hokkaido University, Sapporo
  060-8628, Japan}

\author{Yasuro Funaki} \affiliation{Nishina Center for
  Accelerator-Based Science, The Institute of Physical and Chemical
  Research (RIKEN), Wako 351-0198, Japan}

\author{Hisashi Horiuchi} \affiliation {Research Center for Nuclear
  Physics (RCNP), Osaka University, Osaka 567-0047, Japan}
\affiliation {International Institute for Advanced Studies, Kizugawa
  619-0225, Japan}

\author{\mbox{Gerd R\"{o}pke}} \affiliation{Institut f\"{u}r Physik,
  Universit\"{a}t Rostock, D-18051 Rostock, Germany}

\author{Peter Schuck} \affiliation{Institut de Physique Nucl\'{e}aire,
  Universit\'e Paris-Sud, IN2P3-CNRS, UMR 8608, F-91406, Orsay,
  France} \affiliation{Laboratoire de Physique et Mod\'elisation des
  Milieux Condens\'es, CNRS-UMR 5493, F-38042 Grenoble Cedex 9,
  France}

\author{Akihiro Tohsaki} \affiliation{Research Center for Nuclear
  Physics (RCNP), Osaka University, Osaka 567-0047, Japan}

\author{Chang Xu} \affiliation{School of Physics and Key Laboratory of
  Modern Acoustics, Institute of Acoustics, Nanjing University,
  Nanjing 210093, China}

\author{Taiichi Yamada} \affiliation{Laboratory of Physics, Kanto
  Gakuin University, Yokohama 236-8501, Japan}

\date{\today}
 
\begin{abstract}
  The nonlocalized aspect of clustering, which is a new concept for
  self-conjugate nuclei, is extended for the investigation of the
  $N\not= Z$ nucleus ${}^{9}$Be.  A modified version of the THSR
  (Tohsaki-Horiuchi-Schuck-R\"{o}pke) wave function is introduced with
  a new phase factor. It is found that the constructed negative-parity
  THSR wave function is very suitable for describing the cluster
  states of ${}^{9}$Be. Namely the nonlocalized clustering is shown to
  prevail in ${}^{9}$Be. The calculated binding energy and radius of
  ${}^{9}$Be are consistent with calculations in other models and with
  experimental values. The squared overlaps between the single THSR
  wave function and the Brink+GCM (Generator Coordinate Method) wave
  function for the $3/2^{-}$ rotational band of ${}^{9}$Be are found
  to be near 96\%. Furthermore, by showing the density distribution of
  the ground state of ${}^{9}$Be, the $\pi$-orbit structure is
  naturally reproduced by using this THSR wave function.
\end{abstract}

\pacs{21.60.Gx, 27.20.+n}

\maketitle

\section{Introduction}
The clustering phenomena is one of the fundamental problems in nuclear
physics. Despite of its long history since the discovery of the
$\alpha$-cluster, the clustering structure in nuclei is still under
active investigation \cite{Tohsaki2001, Funaki2002, Yamada2005,
  Zhou2013, Xu2006, Yren2012}. To describe the $\alpha$-cluster
condensation in self-conjugate nuclei ${}^{12}$C and ${}^{16}$O, the
Tohsaki-Horiuchi-Schuck-R\"{o}pke (THSR) wave function was proposed
and provided a successful treatment of the famous Hoyle ($0_{2}^{+}$)
state in ${}^{12}$C \cite{Tohsaki2001, Yamada2005}. The THSR wave
function has been applied to different aggregates of $\alpha$-clusters
including ${}^{8}$Be, ${}^{12}$C and ${}^{16}$O \cite{Funaki2002,
  Tohsaki2001}.  It is also extended to study systems composed of
general clusters such as ${}^{20}$Ne treated as a combination of an
$\alpha$-cluster and ${}^{16}$O \cite{Zhou2012, Zhou2013}.  In the
study of inversion-doublet-band states of ${}^{20}$Ne, the THSR wave
function, which was originally introduced to describe gas-like states,
was shown to be also very suitable for the study of non gas-like
cluster type of states.

The importance of the THSR wave function lies in the fact that the
RGM/GCM (Resonating Group Method/Generator Coordinate Method) wave
functions of both gas-like and non-gaslike states are almost 100\%
equivalent to single THSR wave functions.  Therefore the single THSR
wave function grasps the physical properties of the state. The most
important property is the nonlocalized character of clustering
\cite{Zhou2013}.  On the other hand, superposing many Brink wave
functions in the GCM approach describes non-localized clustering
because the Brink-GCM wave function is almost 100\% equivalent to a
single THSR wave function.

Following the success in describing self-conjugate nuclei, a natural
question is whether we can apply the THSR wave function to general
nuclei which consist not only of $\alpha$-clusters but have also extra
nucleons, such as ${}^{13}$C studied with the interaction of
$\alpha$-condensation and an extra neutron \cite{Tohsaki2008}.
Therefore, it would be promising to extend the THSR wave function to
$N\not=Z$ nuclei.

The nucleus ${}^{9}$Be is a typical $N\not=Z$ nucleus with both
$\alpha+\alpha+n$ cluster structure and also nuclear molecular orbits,
which is most suitable for the extension of the THSR wave
function. This study also belongs to our project which tries to
understand any clustering phenomena from our new point of view. The
nucleus ${}^{9}$Be has been studied with the antisymmetrized molecular
dynamics free from cluster assumptions \cite{Enyo1995}. It is also
investigated with the nuclear molecular orbit (MO) model in which the
extra neutron occupies nuclear molecular orbits \cite{Okabe1977,
  Oertzen1996, Itagaki2000}. In the nuclear molecular orbit model, the
wave function of the extra neutron is assumed to be a linear
combination of cluster orbitals and provides a successful description
of ${}^{9}$Be.  However, the dynamics of the clusters is not
explicitly given in these calculations. To have a clear view of the
nonlocalized clustering dynamics in the ${}^{9}$Be nucleus, we need a
new wave function in which this characteristic is intrinsically
included. In the present work, ${}^{9}$Be is investigated with a new
picture in which two $\alpha$-clusters and an extra nucleon are
performing nonlocalized motion. A modified version of the THSR wave
function is proposed for the description of this structure.

The outline of this paper is as follows. In Section
\ref{sec:waveFunction} we formulate the THSR wave function with
intrinsic parity for ${}^{9}$Be.  Then in Section \ref{sec:results} we
give the results of the calculation for $3/2^{-}$ rotational band of
${}^{9}$Be and analyze the structure of the ground state obtained from
the THSR wave function. The last Section \ref{sec:conclusion} contains
the conclusions.

\section{Formulation of THSR Wave Function for ${}^{9}$B\MakeLowercase{e}}
\label{sec:waveFunction}
The THSR wave function of ${}^{9}$Be is constructed with creation
operators as
\begin{equation}
\label{eq:totalwf}
  \left| \Phi \right\rangle
  =(C_{\alpha}^{\dagger})^{2}c_{n}^{\dagger}\left| {\bf  \rm vac} \right\rangle,
\end{equation}

where $C_{\alpha}^{\dagger}$ and $c_{n}^{\dagger}$ are creation
operators of $\alpha$-particle and neutron, respectively. Here we take
the same $\alpha$-creator $C_{\alpha}^{\dagger}$ as used in previous
deformed THSR wave functions \cite{Funaki2002},
\begin{equation}
  \label{eq:alphaCreator}
  \begin{split}
  C_{\alpha}^{\dagger}=\int &d\mathbf{R}
    \exp (-\frac{R_{x}^{2}}{\beta_{\alpha,xy}^{2}}
    -\frac{R_{y}^{2}}{\beta_{\alpha,xy}^{2}}
    -\frac{R_{z}^{2}}{\beta_{\alpha,z}^{2}})\int d^{3}\mathbf{r}_{1}
      \cdots d^{3}\mathbf{r}_{4}    \\
  &\times \psi(\mathbf{r}_{1}-\mathbf{R})
      a_{\sigma_{1},\tau_{1}}^{\dagger}(\mathbf{r}_{1})
    \cdots \psi(\mathbf{r}_{4}-\mathbf{R})
      a_{\sigma_{4},\tau_{4}}^{\dagger}(\mathbf{r}_{4}),
  \end{split}
\end{equation}

where $\mathbf{R}$ is the generate coordinate of the $\alpha$-cluster,
$\mathbf{r}_i$ is the position of the $i$th nucleon, and the
$a_{\sigma,\tau}^{\dagger}(\mathbf{r}_i)$ is the creation operator of
the $i$th nucleon with spin $\sigma$ and isospin $\tau$ at position
$\mathbf{r}_i$. $\psi(\mathbf{r}) =(\pi b^{2})^{-3/4}
\exp(-r^{2}/2b^{2})$ is the single-nucleon harmonic oscillator shell
model wave function representing one of the four nucleons forming the
alpha cluster, where $b$ is a size parameter. $\beta_{\alpha,xy}$ and
$\beta_{\alpha,z}$ are parameters for the nonlocalized motion of two
$\alpha$-clusters. The subsystem of two $\alpha$-clusters is supposed
to have rotational symmetry around the $z$-axis, so the same size
parameters $\beta_{\alpha,xy}$ are used in $x$ and $y$ direction.  For
the extra neutron, we use a creation operator $c_{n}^{\dagger}$ which
is similar to $C_{\alpha}^{\dagger}$ but has a new introduced phase
factor,
\begin{equation}
  \label{eq:extraCreator}
  \begin{split}
  c_{n}^{\dagger}=\int &d^{3}\mathbf{R}_{n}
    \exp (-\frac{R_{n,x}^{2}}{\beta_{n,xy}^{2}}
          -\frac{R_{n,y}^{2}}{\beta_{n,xy}^{2}}
          -\frac{R_{n,z}^{2}}{\beta_{n,z}^{2}})
    e^{im\phi_{\mathbf{R}_{n}}}\int d^{3}\mathbf{r}_n    \\
  &\times (\pi b^{2})^{-3/4}
  e^{-\frac{(\mathbf{r}_n-\mathbf{R}_{n})^{2}}{2b^{2}}}
    a_{\uparrow,n}^{\dagger}(\mathbf{r}_n),
  \end{split}
\end{equation}
where $\mathbf{R}_{n}$ is the generate coordinate of the extra
neutron, $\mathbf{r}_n$ is the position of the extra neutron,
$a_{\sigma,\tau}^{\dagger}(\mathbf{r}_n)$ is the creation operator of
the extra neutron with spin up at position $\mathbf{r}_n$, and
$\phi_{\mathbf{R}_{n}}$ is the azimuthal angle in spherical
coordinates $(R_{\mathbf{R}_{n}}, \theta_{\mathbf{R}_{n}},
\phi_{\mathbf{R}_{n}})$ of $\mathbf{R}_{n}$. In this creation
operator, the same size parameter $b$ of Gaussian is used as in
Eq.~(\ref{eq:alphaCreator}).  $\beta_{n,xy}$ and $\beta_{n,z}$ are
parameters for the nonlocalized motion of the extra neutron.

To illustrate the detailed structure of our wave function and prove
its negative parity, we can rewrite the THSR wave function of
${}^{9}$Be with $m=\pm 1$ in the form of,
\begin{equation}
  \label{eq:singleFunction}
  \langle \mathbf{r}_1,\sigma_1,\tau_1,\dots,\mathbf{r}_n,\sigma_n,\tau_n
  \left| \Phi \right\rangle \propto \mathcal{A}
     \left[ F(\mathbf{X}_1,\mathbf{X}_2, \mathbf{r}_n)
       \phi(\alpha_1)\phi(\alpha_2) 
     \right]
\end{equation}
where $\mathcal{A}$ is an antisymmetrizer, $\phi(\alpha_1)$ and
$\phi(\alpha_2)$ are internal $\alpha$ wave functions as in
Ref.~\cite{Tohsaki2001}, $\mathbf{X}_1$ and $\mathbf{X}_2$ are center
of mass coordinates of the two $\alpha$-clusters, and the form of
function $F(\mathbf{X}_1,\mathbf{X}_2, \mathbf{r}_n)$ is,
\begin{equation}
  \label{eq:singleFunctionF}
  \begin{split}
F(\mathbf{X}_1,\mathbf{X}_2, \mathbf{r}_n) =  
  &\exp \left\{-\frac{X_{1,x}^2+X_{1,y}^2+X_{2,x}^2+X_{2,y}^2}{C_{\alpha,xy}^2}
                     -\frac{X_{1,z}^2+X_{2,z}^2}{C_{\alpha,z}^2}
        \right\}  \\
  &\times \exp \left\{-\frac{r_{n,x}^2+r_{n,y}^2}{C_{n,xy}^2}
                     -\frac{r_{n,z}^2}{C_{n,z}^2}
               \right\}\\
  &\times [I_0(\frac{r_{n,x}^2+r_{n,y}^2}{A^2})
           +I_1(\frac{r_{n,x}^2+r_{n,y}^2}{A^2})]\\
  &\times (r_{n,x}^2+r_{n,y}^2)^{1/2} e^{\pm i \phi_{r_n}}
  \end{split}
\end{equation}          
where $I_0$ and $I_1$ are the modified Bessel functions of the first
kind, $\phi_{r_n}$ is the azimuthal angle of $\mathbf{r}_n$, and the
denominators are, $C_{\alpha,xy}^2 = b^2/2+\beta_{\alpha,xy}^2$,~
$C_{\alpha,z}^2 = b^2/2+\beta_{\alpha,z}^2$, ~ $C_{n,xy}^2 =
(8b^4+4b^2\beta_{n,xy}^2)/(4b^2+\beta_{n,xy}^2)$,~ $C_{n,z}^2 = 2
b^2+\beta_{n,z}^2$, ~and~ $A^2 =4b^2+ 8b^4/\beta_{n,xy}^2$.  The
detailed derivation of Eq.~(\ref{eq:singleFunction}) and
Eq.~(\ref{eq:singleFunctionF}) can be found in the appendix.

The wave function in Eq.~(\ref{eq:singleFunction}) gives a clear view
of the dynamics of motions inside nucleus ${}^9$Be as a single
function without superposition.  It consists of two parts, the
internal motion of nucleons inside $\alpha$-clusters $\phi(\alpha_1)$
and $\phi(\alpha_2)$, and the center-of-mass motions of clusters and
the extra neutron described by $F(\mathbf{X}_1,\mathbf{X}_2,
\mathbf{r}_n)$.  The negative parity of our THSR wave function with
$m=\pm 1$ can be clearly obtained with Eq.~(\ref{eq:singleFunction})
and Eq.~(\ref{eq:singleFunctionF}).  The Gaussians and the modified
Bessel functions in function $F$ will not change under the inversion
of space. For the last phase factor $e^{i \phi_{r_n}}$ in function
$F$, we have ${\hat P}_r e^{\pm i \phi_{r_n}}=e^{\pm i
  (\pi+\phi_{r_n})}=-e^{\pm i \phi_{r_n}}$, which results in a
negative parity for the function $F(\mathbf{X}_1,\mathbf{X}_2,
\mathbf{r}_n)$. Considering that internal wave functions
$\phi(\alpha_1)$ and $\phi(\alpha_2)$ have positive parity, thus the
negative parity of the THSR wave function with $m=\pm 1$ is demonstrated.

We give another proof of the negative parity of our ${}^9$Be wave
function without executing the integrations over $\mathbf{R}$ and
$\mathbf{R}_n$: The spatial wave function of the extra neutron
$\Phi_{n}(\mathbf{r}_n)$ can be written as,
 \begin{equation}
   \label{eq:extraWaveFunction}
   \begin{split}
 \Phi_{n}(\mathbf{r}_n)=\int &d^{3}\mathbf{R}_n
     \exp (-\frac{R_{n,x}^{2}}{\beta_{n,xy}^{2}}
     -\frac{R_{n,y}^{2}}{\beta_{n,xy}^{2}}
     -\frac{R_{n,z}^{2}}{\beta_{n,z}^{2}})e^{im\phi_{\mathbf{R}_n}}\\
         &\times (\pi b^{2})^{-3/4}e^{-\frac{(\mathbf{r}_n-\mathbf{R}_n)^{2}}{2b^{2}}}.
   \end{split}
 \end{equation}
 When we change the integration variables $(R_{n,x}, R_{n,y},
 R_{n,z})$ to $(-R_{n,x}, -R_{n,y}, -R_{n,z})$, in the integral
 representation of $\Phi_n(- r_n)$ by
 Eq.~(\ref{eq:extraWaveFunction}), we obtain $\Phi_n(- r_n) =
 -\Phi_n(r_n)$.  It is because the azimuthal angle of $(-R_{n,x},
 -R_{n,y}, -R_{n,z})$ is $(\pi+\phi_{\mathbf{R}_n})$ and
 $\exp(i(\pi+\phi_{\mathbf{R}_n})) = - \exp(i\phi_{\mathbf{R}_n})$.
 When $m=0$, Eq.~(\ref{eq:extraWaveFunction}) is a standard THSR wave
 function and has a positive parity as already known before in
 Ref.~\cite{Zhou2013}.  So the total parity of ${}^{9}$Be is now
 determined by $m$,
\begin{equation}
  \pi=\pi_{\alpha}^{(1)}\times\pi_{\alpha}^{(2)}\times\pi_{n}=
  \begin{cases}
    +& (m=0)\\
    -& (m=\pm1).
  \end{cases}
\end{equation}

From Eq.~(\ref{eq:singleFunctionF}) we can also see that the $z$
component of orbital angular momentum $l_{z,n}=m$ is a good quantum
number for the extra neutron. Because of the rotational symmetry of
the two-$\alpha$-cluster subsystem about $z$-axis, we have
$l_{z,\alpha}=0$ for $\alpha$ clusters. Thus the $z$-component of the
orbital angular momentum $l_{z}$ of total system is
\begin{equation}
\label{eq:lz}
  l_{z}=l_{z,\alpha}^{(1)}+l_{z,\alpha}^{(2)}+l_{z,n}=m.
\end{equation}

In order to eliminate effects from spurious center-of-mass (c.o.m.)
motion, the c.o.m. part of $\left| \Phi \right\rangle$ is projected
onto a $(0s)$ state \cite{Okabe1977},
\begin{equation}
\left| \Psi \right\rangle=\left| (0s)\text{c.o.m.} \right\rangle\rangle
    \left\langle\langle (0s)\text{c.o.m.}  | \Phi \right\rangle.
\end{equation} 
Here $(0s)$ represents the wave function of the c.o.m. coordinate
$\mathbf{X}_{G}$, which is the ground state of harmonic
oscillator. For the THSR wave function of ${}^9$Be, this projection
can be accomplished simply by the transformation of coordinates
$\mathbf{r}_{i}$ in $\left| \Phi \right\rangle$ as
\begin{equation}
 \mathbf{r}_{i} \rightarrow \mathbf{r}_{i} -\mathbf{X}_{G}.
\end{equation}
Then the spurious center-of-mass motion in the wave function can be
separated and eliminated analytically.

We also apply the angular-momentum projection technique
$\hat{P}_{MK}^{J}\left| \Psi \right\rangle$ to restore the rotational
symmetry \cite{Schuck1980},
\begin{equation}\label{eq:ap}
  \begin{split}
\left| \Psi^{JM} \right\rangle&=\hat{P}_{MK}^{J}\left| \Psi \right\rangle\\
    &=\frac{2J+1}{8\pi^{2}}\int d \Omega D^{J*}_{MK}(\Omega)\hat R (\Omega)
    \left| \Psi \right\rangle,
  \end{split}
\end{equation}
 where $J$ is the total angular momentum of ${}^9$Be.

The Hamiltonian of ${}^{9}$Be system can be written as
\begin{equation}\label{hamiltonian}
  H=\sum_{i=1}^{9} T_i-T_{c.m.} +\sum_{i<j}^{9}V^N_{ij}
    +\sum_{i<j}^{9}V^C_{ij} +\sum_{i<j}^{9}V^{ls}_{ij},
\end{equation}
where $T_{c.m.}$ is the kinetic energy of the center-of-mass
motion. Volkov No. 2 \cite{Volkov1965} is used as the central force of
nucleon-nucleon potential,
\begin{equation}\label{vn}
  V^N_{ij}=\{V_1 e^{-\alpha_1 r^2_{ij}}-V_2 e^{-\alpha_2 r^2_{ij}}\}
  \{ W - M \hat P_\sigma \hat P_\tau \ + B \hat P_{\sigma} - H \hat P_{\tau}\},
\end{equation}
where $M=0.6$, $W=0.4$ and $B=H=0.125$. Other parameters are
$V_{1}=-60.650$ MeV, $V_{2}=61.140$ MeV,  $\alpha_{1}=0.309$
fm${}^{-2}$, and $\alpha_{2}=0.980$ fm${}^{-2}$.

In traditional THSR calculations of 4-n nuclei, the spin-orbit
interaction cancels out and can be safely neglected.  However, for the
THSR calculation of ${}^9$Be, the spin-orbit interaction plays a key
role because of the existence of extra neutron.  The G3RS (Gaussian
soft core potential with three ranges) term \cite{Yamaguchi1979},
which is a two-body type interaction, is taken as the spin-orbit
interaction,
\begin{equation}\label{vc}
V^{ls}_{ij}=V^{ls}_0\{ 
           e^{-\alpha_1 r^2_{ij}}-e^{-\alpha_2 r^2_{ij}}
           \} \mathbf{L}\cdot\mathbf{S} \hat{P}_{31},
\end{equation} 
where $\hat P_{31}$ projects the two-body system into triplet odd
state.  Parameters in $V^{ls}_{ij}$ are taken from
Ref.~\cite{Okabe1979} with $V_{0}^{ls}$=2000 MeV, $\alpha_{1}$=5.00
fm${}^{-2}$, and $\alpha_{2}$=2.778 fm${}^{-2}$.

\section{Results and Discussions}
\label{sec:results}
Here we calculated the ground state properties of ${}^{9}$Be, which is
a stable bound state and the only state that exists below the
$\alpha+\alpha+n$ threshold.  The excited states $5/2^{-}$ and
$7/2^{-}$, which belong to the rotational band of the $3/2^{-}$ ground
state, are also calculated.

For the intrinsic wave function, $m=1$ is taken in the phase factor
$e^{im\phi_{\mathbf{R}_{n}}}$ in Eq.~(\ref{eq:extraCreator}) to ensure
the negative parity in these states.  Because the $l_{z}=m$ is a good
quantum number for the intrinsic wave function as shown in
Eq.~(\ref{eq:lz}), we can write the $z$-component of total angular
momentum as $K=l_z\pm1/2$ for the parallel and anti-parallel coupling
of spin and the orbital angular momentum.  As we will show later,
because of the existence of spin-orbital interaction, the intrinsic
wave function with $K=3/2$ has a much lower energy than $K=1/2$.  In
the following calculations, $K=3/2$ is taken in the angular momentum
projection operators.

It should be noted here that $m$ and $K$ are parameters related to the
intrinsic wave function only, while parameters $J$ and $M$ describe
the state after angular momentum projection.  The parameter $M$ can be
chosen freely.

The binding energy of ${}^{9}$Be can be obtained by a variational
calculation with respect to parameters $b, \beta_{\alpha,xy},
\beta_{\alpha,z}, \beta_{n,xy},$ and $\beta_{n,z}$ as,
\begin{equation}
  \label{eq:bindingEnergy}
  E(b,\beta_{\alpha,xy},\beta_{\alpha,z},\beta_{n,xy},\beta_{n,z})
    =\frac{\langle \Psi^{JM} |\hat H|\Psi^{JM} \rangle}
    {\langle \Psi^{JM} |\Psi^{JM} \rangle}.   
\end{equation}
The Monte Carlo method is used for the numerical integration of Euler
angle $\Omega$ in the angular momentum projection and coordinates
$\{\mathbf{R}, \mathbf{R}_n\}$ in creation operators.

In our calculations, we treat $b$ as a variational parameter and get
as the optimum value $b=1.35$ fm, which is the same as that obtained
for ${}^{8}$Be \cite{Funaki2002}. This shows that the size of each
$\alpha$ clusters in ${}^{9}$Be is similar to the ones in
${}^{8}$Be. This also shows a relatively weak influence of the extra
neutron on the $\alpha$-clusters. The remaining optimum parameters in
the THSR wave function are $\beta_{\alpha,xy}=0.1$ fm,
$\beta_{\alpha,z}=4.2$ fm, $ \beta_{n,xy}=2.5$ fm, and
$\beta_{n,z}=2.8$ fm.

For the ground state of ${}^9$Be, we get an energy of -55.4 MeV with
the spin-orbit term of -2.2 MeV with the THSR wave function. This
negative value shows that, with $K=3/2$, the system energy is lowered
by the spin-orbit interaction. As a comparison, with the same
parameters and $K=1/2$, we get a much higher energy of -50.3 MeV and a
positive spin-orbit term of 2.2 MeV. Thus, the parallel coupling
structure in the intrinsic wave function is preferred as we discussed
above.

In Table \ref{table:overlap} we show the calculated results of the
$3/2^{-}$ rotational band. Binding energies calculated with both THSR
wave function and the Brink+GCM technique are included. The Brink+GCM
results are calculated with the same interactions as used in the THSR
calculations.  We use 72 Brink wave functions of ${}^9$Be with 6
different $\alpha$-$\alpha$ distances and 12 different positions of
the extra neutron as the bases in the Brink+GCM calculation.  For all
three states in the rotational band, the calculated results from THSR
wave function agree well with the Brink+GCM results. The excitation
energies calculated with both THSR and Brink+GCM method are consistent
with the experimental values. The differences between theoretical
results and experimental results in binding energy is due to the
choice of interactions. We also show the squared overlap between the
THSR wave function and the Brink+GCM wave function. The calculated
squared overlap for the ground state state is about 96\% and the
results for the excited states are similar. As the Brink+GCM wave
function is generally considered as the exact wave function of the
system, these overlaps show that the single THSR wave function
provides a good description for these states.

\begin{table*}[htbp]
  \begin{center}
    \caption{\label{table:overlap}Calculation results of the $3/2^{-}$
      rotational band. G.S. denotes the ground state and E.S. denotes
      the excited state. $E^{\text{THSR}}$ and $E^{\text{GCM}}$ are
      calculated binding energy with the THSR wave function and the
      Brink+GCM wave function respectively. $E^{\text{Exp}}$ is the
      experimental result. Values in parentheses are corresponding
      excitation energies.
      $|\langle\Psi^{\text{THSR}}|\Psi^{\text{GCM}}\rangle|^{2}$ is
      the squared overlap between the THSR wave function and the
      Brink+GCM wave function. All units of energies are MeV.}
  \begin{tabular}{l c c c c}
    \hline
    \hline
State &$E^{\text{THSR}}$  &$E^{\text{GCM}}$ 
&$E^{\text{Exp}}$\cite{Tilley2004, Audi2012}
            &$|\langle\Psi^{\text{THSR}}|\Psi^{\text{GCM}}\rangle|^{2}$\\
    \hline
$7/2^-$(E.S.) &48.6 (6.8)  &49.4 (7.0) &51.8 (6.4) &0.93\\
$5/2^-$(E.S.) &53.0 (2.4)  &53.8 (2.6) &55.8 (2.4) &0.95\\
$3/2^-$(G.S.) &55.4        &56.4       &58.2       &0.96\\
    \hline
    \hline
  \end{tabular}
  \end{center}
\end{table*}

The  root-mean-square (RMS) radius 
is also calculated for the ground state of ${}^{9}$Be
with,
\begin{equation}
  \label{eq:rms}
  r_{RMS}=
  \sqrt{
    \frac{
      \left\langle \Psi^{JM} \right|
      \frac{1}{9}\sum_{i=1}^{9}
      (\mathbf{r}_{i}-\mathbf{X}_{G})^{2}
      \left| \Psi^{JM} \right\rangle
    }{
      \left\langle \Psi^{JM} |
        \Psi^{JM} \right\rangle}}.
\end{equation}
With all parameters variationally optimized, the THSR wave function
gives a point-RMS radius of 2.55 fm for the ground state, which agrees well
with the experimental value 2.45 fm \cite{Tanihata1985}.

\begin{figure}[htbp]
  \centering
  \includegraphics[width=0.45\textwidth]{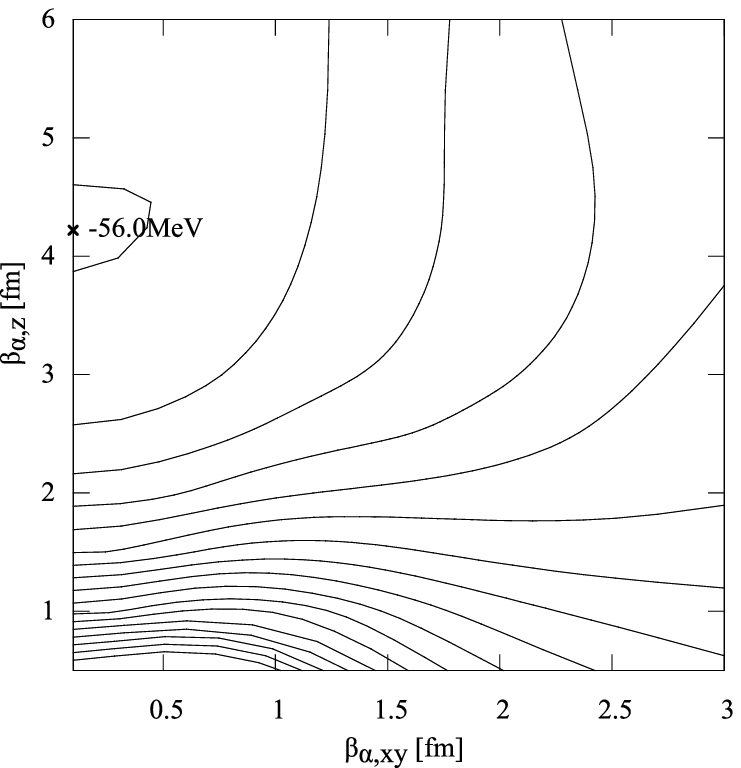}
  \includegraphics[width=0.45\textwidth]{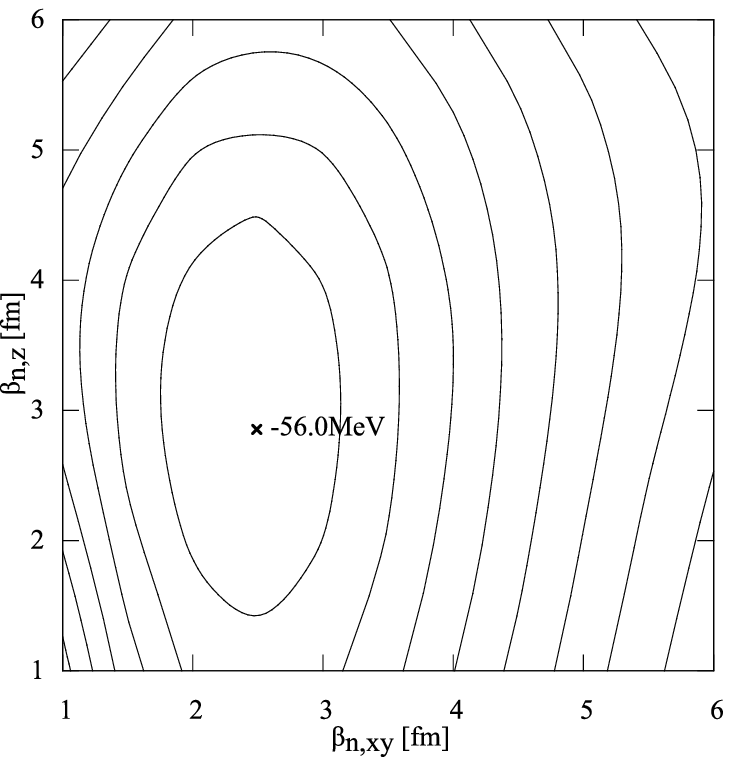}
  \caption{\label{fig:varsize}Contour maps of binding energy surface
    with the $\beta$ parameters in the THSR wave function. Left part
    (a) is the contour map of parameters $\beta_{\alpha,xy}$ and
    $\beta_{\alpha,z}$ and the right part (b) is the contour map of
    parameters $\beta_{n,xy}$ and $\beta_{n,z}$. The optimum value is
    marked on each map labeled with coordinates.  Parameter $b$ is
    taken as the variational optimum value $b=1.35$ fm}.
\end{figure}

\begin{figure}[htbp]
  \centering
  \includegraphics[width=0.45\textwidth]{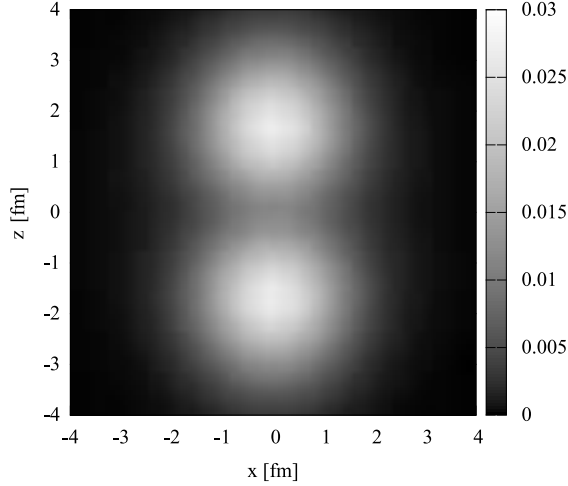}
  \caption{\label{fig:densityTotal}Density distribution of the
    intrinsic ground state of ${}^{9}$Be. The gray scale of each point
    in the figure stands for the nucleon density on $x-z$ plane of the
    $y=0$ cross section. The unit of the density is fm$^{-3}$.  }
\end{figure}

Fig.~\ref{fig:varsize} (a) and (b) are contour maps of the ground
state binding energy surface. The optimum parameters for the ground
state are labeled in the map.  The very large difference between
$\beta_{\alpha,xy}$ and $\beta_{\alpha,z}$ indicates a long prolate
shape of $\alpha$ cluster distribution which is surrounded by the less
deformed distribution of the extra neutron. This configuration
indicates a structure of nuclear molecular orbit for the ground state
of ${}^{9}$Be.  In order to illustrate this structure in detail, the
density distribution of ${}^{9}$Be is calculated as the expectation
value of the density operator,
\begin{equation}
  \label{eq:totalDensity}
  \rho(\mathbf{r'})=\left\langle \Psi \right|
    \frac{1}{9}\sum_{i=1}^{9}
    \delta(\mathbf{r}_{i}-\mathbf{X}_{G}-\mathbf{r'})
    \left| \Psi \right\rangle.
\end{equation}
where $\Psi$ is the normalized intrinsic THSR wave function of
${}^9$Be.  Fig.~\ref{fig:densityTotal} shows the density distribution
in the $y=0$ cross section. Similar to the case of ${}^{8}$Be, a clear
structure of two $\alpha$ clusters is displayed. Due to the Pauli
blocking effect, the two $\alpha$ clusters can not get too close to
each other and a neck structure appears.  The distance between two
$\alpha$-clusters in ${}^{9}$Be is about 3.4 fm while the
inter-cluster distance for the ground state of ${}^{8}$Be is about 4.6
fm. This shows a more compact structure and a stronger binding effect
of two $\alpha$-clusters in ${}^{9}$Be because of the existence of the
extra nucleon.

\begin{figure}[htbp]
  \centering
  \includegraphics[width=0.45\textwidth]{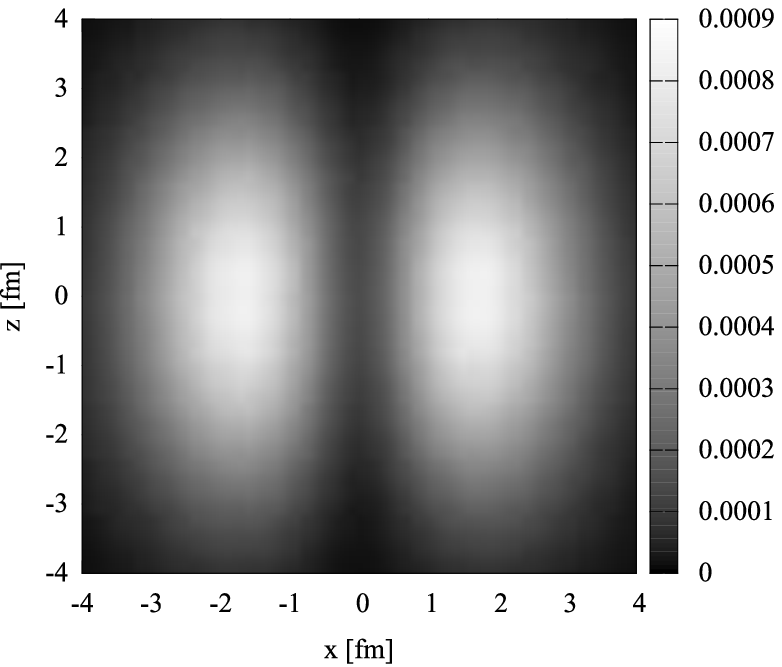}
  \includegraphics[width=0.45\textwidth]{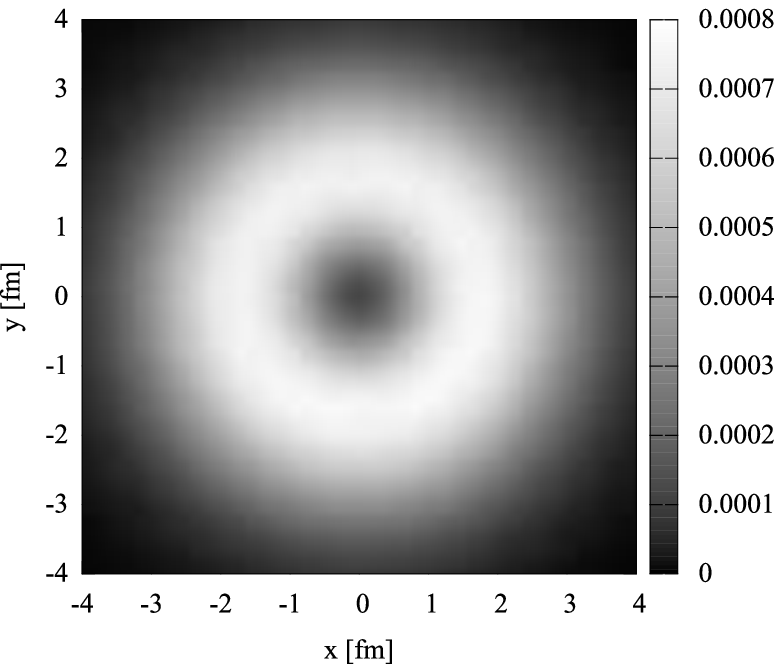}
  \caption{\label{fig:densityExtra}Density distribution
    $\rho(\mathbf{r}'_n)$ of the extra neutron of the intrinsic ground
    state of ${}^9$Be.  The gray scale of each point in left part (a)
    of the figure stands for the nucleon density on $x-z$ plane of the
    $y=0$ cross section.  The gray scale of each point in right part
    (b) of the figure stands for the nucleon density on $x-y$ plane of
    the $z=0$ cross section.  The unit of the density is fm$^{-3}$.  }

\end{figure}

To get a clear view of the binding effect of the extra neutron and the
structure of the ground state, we also calculate the density
distribution $\rho(\mathbf{r}'_n)$ of the extra nucleon. The intrinsic
wave function $\Psi$ can be written in the form of,
\begin{equation}
  \Psi = \text{C} \hat A[\Phi^{ \text{THSR}}(2 \alpha) \phi_n(\mathbf{r})],
\end{equation}
where $\hat A$ is the antisymmetrizer, and C is the normalization
constant.  Then we can define the density distribution
$\rho(\mathbf{r}'_n)$ of the extra neutron as
\begin{equation}
  \rho(\mathbf{r}'_n) = \sqrt{9} \langle \Phi^{ \text{THSR}}(2 \alpha)
  \phi_n(\mathbf{r})  |
  \delta(\mathbf{r} - \mathbf{X}_G - \mathbf{r}'_n) | \Psi \rangle,
\end{equation}
where $\sqrt{9}$ comes from the normalization constant
\cite{Horiuchi1977}.  As shown in Fig.~\ref{fig:densityExtra}, a
distribution result which consists of two parts is displayed on the
$y=0$ cross section and a ring style distribution can be seen on the
$z=0$ cross section. This distribution, in which the extra neutron
cannot stay along the $z$-axis, originates from the restriction of
rotational symmetry by the phase factor $e^{i
  \phi_{\mathbf{R}_n}}$. However, this restriction is reasonable
because similar distribution has been given by previous GCM
calculations \cite{Okabe1977,Okabe1979}. The distribution of the extra
nucleon spreads more than 6 fm in $z$ direction, which is about double
the size of each $\alpha$ cluster. It also has overlaps with
distributions of both $\alpha$ clusters, which is known as the $\pi$
orbit in nuclear Molecular Orbit model. This is interesting because,
at variance with previous works in Ref.~\cite{Oertzen1996} and
Ref.~\cite{Itagaki2000}, no molecular orbit is presumed in our wave
function.  In the THSR wave function, the extra nucleon is only
assumed to make a nonlocalized motion inside the nucleus. The
$\pi$-orbit emerges naturally from the antisymmetrization which
cancels out nonphysical distributions.  This reproduction of nuclear
molecular orbit structure provides another support for our extension
of the nonlocalized clustering concept to ${}^{9}$Be.

\begin{table*}[htbp]
  \begin{center}
    \caption{\label{table:energy}Comparison of results form the THSR
      wave function and the nuclear Molecular Orbit model. $E$ is the
      binding energy in MeV. ``THSR'' denotes the result calculated
      with THSR wave function.  ``MO'' denotes the result of Molecular
      Orbit model and ``MO+GCM'' is the result of Molecular Orbit
      model plus GCM technique. Parameter $b=1.46$ fm as used in
      Ref. \cite{Itagaki2000}. Other parameters are variationally
      optimized.}
  \begin{tabular}{c c c}
    \hline
    \hline
Model    &$b$ (fm) &$E(3/2^-)$ (MeV)\\
    \hline
THSR                         &1.46 &54.7\\
MO \cite{Itagaki2000}        &1.46 &54.8\\
MO + GCM \cite{Itagaki2000}  &1.46 &56.1\\
    \hline
    \hline
  \end{tabular}
  \end{center}
\end{table*}

To compare the THSR wave function with the nuclear molecular orbit
model, we use $b=1.46$ fm which is the same value as in
Ref.~\cite{Itagaki2000}. The calculated binding energy of the ground
state $3/2^{-}$ with different models but same interaction and
parameter $b$ are listed in Table \ref{table:energy}. With the THSR
wave function, we get a value of -54.7 MeV for the binding energy of
the ground state, which is almost the same as the result of the
Molecular Orbit (MO) Model without GCM technique in
Ref.~\cite{Itagaki2000}. This agreement shows that the motion of the
valence neutron in ${}^{9}$Be is well treated with the THSR wave
function. Comparison with the result of MO+GCM method in
Ref.~\cite{Itagaki2000} shows that our result is about 1.3 MeV
higher. This is acceptable because the results of MO+GCM method should
be compared with results of THSR+GCM model rather than with those from
a single THSR wave function.

We consider that both $3/2^-$ ground state and other states such as
$1/2^+$ state will be well described by single THSR wave functions.
This paper has shown that really a single THSR wave function well
describes the $3/2^-$ ground state. In our near-future paper we will
study the $1/2^+$ state.

\section{Conclusion}
\label{sec:conclusion}
We extended the nonlocalized clustering concept inherent to the THSR
wave function to the $N\not=Z$ nucleus ${}^{9}$Be, in which the
$\alpha$-clusters and extra neutron make nonlocalized motion inside
the nucleus.  We introduce a modified version of the THSR wave
function that includes a creation operator of the extra neutron.  With
the introduced phase factor $e^{im\phi_{\mathbf{R}}}$, our wave
function has intrinsic negative parity for $m=\pm1$. Binding energies
are calculated for the $3/2^{-}$ rotational band head of ${}^{9}$Be by
the variational method.  The calculated binding energy from the THSR
wave function fits well with the Brink+GCM results. The excitation
energies of two excited states are also reasonable compared with the
experimental results. The squared overlap between the THSR wave
function and the Brink+GCM wave function are found to be close to
96\%.  This means that the THSR wave function provides a good
description of the $3/2^{-}$ rotational band head of ${}^{9}$Be.  With
the same parameter $b=1.46$ fm, our result for the binding energy of
the ground state is consistent with the Molecular Orbit (MO) model but
higher than in the MO+GCM model.  The calculated RMS radius of the
ground state also agrees well with the experimental value.  By
calculating density distributions of the ground state of ${}^{9}$Be,
the $\pi$-orbit structure is naturally reproduced by the THSR wave
function without ad hoc assumption. The calculation of ${}^{9}$Be
provides support for the extension of the nonlocalized clustering
concept to $N\not=Z$ nuclei. It also shows to possess the flexibility
to describe other structures such as nuclear molecular orbital
structure with the THSR wave function.  Though with our technique we
essentially have not found anything for the ${}^9$Be structure which
was not known before, we think that it is interesting to see that the
THSR wave function also works with adding valence neutrons to the
$\alpha$ particles. This is because it is shown that the geometrical
cluster structures arise only from kinematical reasons which is a new
aspect of cluster physics. Otherwise clusters and extra neutrons are
free in their motion.

\begin{acknowledgments}
  This work is supported by the National Natural Science Foundation of
  China (grant nos 11035001, 11375086, 11105079, 10735010, 10975072,
  11175085 and 11235001), by the 973 National Major State Basic
  Research and Development of China (grant nos 2013CB834400 and
  2010CB327803), by the Research Fund of Doctoral Point (RFDP), grant
  no. 20100091110028, and by the Science and Technology Development
  Fund of Macao under grant no. 068/2011/A.
\end{acknowledgments}

\appendix*
\section{The derivation of the single form of the THSR wave function with $m=1$}
The THSR wave function of ${}^9$Be can be write in $r$ space as
\begin{equation}\label{eq:thsr}
  \begin{split}
  &\langle \mathbf{r}_1,\sigma_1,\tau_1,\dots,\mathbf{r}_n,\sigma_n,\tau_n
   \left| \Phi \right\rangle\\
=& \int d\mathbf{R}_1\int d\mathbf{R}_2\int d\mathbf{R}_n
    \prod_{i=1,2,n}\prod_{k=x,y,z}
    \exp \left\{-\frac{R_{i,k}^{2}}{\beta_{i,k}^{2}} \right\}
    \exp\{i \phi_{\mathbf{R}_n}\}
    \Phi^B(\mathbf{R}_1,\mathbf{R}_2,\mathbf{R}_n) 
  \end{split}
\end{equation}
where $\Phi^B$ is the Brink wave function.  To obtain the single
function form of the THSR wave function, we need to perform the
integration of the generate coordinates $\mathbf{R}_1$, $\mathbf{R}_2$
and $\mathbf{R}_n$ in Eq.~(\ref{eq:thsr}) analytically. Because the
Brink wave function is the antisymmetrization of single nucleon wave
functions, this integration over different generate coordinates can be
performed separately.

The integration over two $\alpha$-cluster generate coordinates
$\mathbf{R}_1$ and $\mathbf{R}_2$ is already given as
$F_\alpha(\mathbf{R}_1, \mathbf{R}_2) \phi(\alpha_1)\phi(\alpha_2)$ in
Ref.~\cite{Tohsaki2001}, where $\Psi(\alpha)$ is the internal
$\alpha$-cluster wave function, and the motion of the center-of-mass
of $\alpha$-clusters is
\begin{equation}
   \begin{split}
 F_\alpha(\mathbf{X}_1,\mathbf{X}_2) =  
   \exp \left\{-\frac{X_{1,x}^2+X_{1,y}^2+X_{2,x}^2+X_{2,y}^2}{C_{\alpha,xy}^2}
                      -\frac{X_{1,z}^2+X_{2,z}^2}{C_{\alpha,z}^2}
         \right\},  \\
   \end{split}
\end{equation}
where $C_{\alpha,xy}^2 = b^2/2+\beta_{\alpha,xy}^2$ and
$C_{\alpha,z}^2  = b^2/2+\beta_{\alpha,z}^2$.

The integration over generate coordinate $\mathbf{R}_n$ for the extra
neutron can be separated into two integration of its different
components.  The first one is the integration over its $z$-component
$R_{n,z}$ as,
\begin{equation}\label{eq:intz}
  \begin{split}
f_{n, z}(r_{n,z})=&
 \int dR_{n,z}
    \exp \left\{-\frac{R_{n,z}^{2}}{\beta_{n,z}^{2}} \right\}
    \exp \left\{-\frac{1}{2 b^2}(r_{n,z}-R_{n,z})^2 \right\}\\
\propto&\exp \left\{-\frac{r_{n,z}^2}{C_{n,z}^2}     \right\}
  \end{split}
\end{equation}
where $C_{n,z}^2       = 2 b^2+\beta_{n,z}^2$.
Another one is the integration over its $x$-component $R_{n,x}$
and $y$-component $R_{n,y}$ as
\begin{equation}
  \begin{split}
f_{n, xy}(r_{n,x},r_{n,y})=
 \int dR_{n,x}dR_{n,y}&
    \exp \left\{-\frac{R_{n,x}^{2}+R_{n,y}^{2}}{\beta_{n,xy}^{2}} \right\}
    e^{i \phi_{\mathbf{R}_n}}\\
&\times\exp \left\{-\frac{1}{2 b^2}
     [(r_{n,x}-R_{n,x})^2+(r_{n,y}-R_{n,y})^2)]\right\}.
  \end{split}
\end{equation}
This integration of $R_{n,x}$ and $R_{n,y}$ can be rewritten in polar coordinate
system as
\begin{equation}
  \begin{split}
f_{n, xy}(r_{n,x}, r_{n,y})=
  \int_0^{2\pi} d\phi_R \int_0^{\infty} \rho_R d\rho_R &
    \exp \left\{-\frac{\rho^{2}_R}{\beta_{n,xy}^{2}} \right\}
    e^{i \phi_R}\\
    &\times \exp \left\{
      -\frac{1}{2 b^2}[\rho^2_R+\rho_r^2+2\rho_R\rho_r\cos(\phi_R-\phi_r)]\right\},   
  \end{split}
\end{equation}
where $\rho_R=(R_{n,x}^{2}+R_{n,y}^{2})^{1/2}$,
$\rho_r=(r_{n,x}^{2}+r_{n,y}^{2})^{1/2}$, and $\phi_R$ and $\phi_r$
are angles of vector $(R_{n,x},R_{n,y})$ and $(r_{n,x},r_{n,y})$ in
polar coordinate system respectively. Obviously, we have
$\phi_R=\phi_{\mathbf{R}_n}$, where $\phi_{\mathbf{R}_n}$ is the
azimuthal angle of $\mathbf{R}_n$ in spherical coordinate system.

As a first step, the integration over radial coordinate $\rho_R$ is
performed as
\begin{equation}\label{eq:intOverRho}
  \begin{split}
  &\int \rho_R d\rho_R
    \exp \left\{-\frac{\rho^{2}_R}{\beta_{n,xy}^{2}} 
      -\frac{1}{2 b^2}[\rho^2_R+\rho_r^2+2\rho_R\rho_r\cos(\phi_R-\phi_r)]\right\}  \\ 
\propto& \exp \left\{-\frac{1}{2 b^2}\rho_r^2 \right\}\\
 &\times
     \left\{ a+\sqrt{\pi}\rho_r \cos(\phi_R-\phi_r) 
              [\text{Erf} (\rho_r \cos(\phi_R-\phi_r)/a)-1]
              \exp(\rho_r^2\cos^2(\phi_R-\phi_r)/a^2) \right\},
  \end{split}
\end{equation}
where $a=\sqrt{2b^2+4b^4/\beta_{xy}^2}$. 

The next step is to integrate over the angle $\phi_R$.  Since
$\int_0^{2\pi}\exp(i\phi_R)d\phi_R=0$, we can safely omit the first
constant $a$ in Eq.~(\ref{eq:intOverRho}) as
\begin{equation}
  \begin{split}
f_{n, xy}(r_{n,x}, r_{n,y})\propto \int_0^{2\pi}d\phi_R  &
         \exp \left\{-\frac{1}{2 b^2}\rho_r^2 + i\phi_R \right\}
           \rho_r \cos(\phi_R-\phi_r) \\
 &\times     
         [\text{Erf} (\rho_r \cos(\phi_R-\phi_r)/a)-1]
         \exp(\rho_r^2\cos^2(\phi_R-\phi_r)/a^2).
\end{split}
\end{equation}
Then the substitution $\phi_R\rightarrow\phi_R-\phi_r+\phi_r$ is
applied to the equation above as
\begin{equation}
  \begin{split}
f_{n, xy}(r_{n,x}, r_{n,y})\propto \int_0^{2\pi}d\phi_R 
&\exp \left\{-\frac{1}{2 b^2}\rho_r^2 + i\phi_r \right\}
           e^{i(\phi_R-\phi_r)}
           (\rho_r \cos(\phi_R-\phi_r))\\
 &\times     
         [\text{Erf} (\rho_r \cos(\phi_R-\phi_r)/a)-1]
         \exp(\rho_r^2\cos^2(\phi_R-\phi_r)/a^2).
\end{split}
\end{equation}
Since $f_{n, xy}(r_{n,x}, r_{n,y})$ is a periodic function of $\phi_R$
with a $2\pi$ period, the integration over $\phi_R$ can be written as
\begin{equation}
  \begin{split}
f_{n, xy}(r_{n,x}, r_{n,y})\propto \int_0^{2\pi}
&\exp \left\{-\frac{1}{2 b^2}\rho_r^2 + i\phi_r \right\}
           e^{i\theta}
           (\rho_r \cos\theta)\\
 &\times     
         [\text{Erf} (\rho_r \cos\theta/a)-1]
         \exp(\rho_r^2\cos^2 \theta/a^2)d\theta,
  \end{split}
\end{equation}
where $\theta=\phi_R-\phi_r$.

Considering the Euler equation $e^{i\theta}=\cos\theta+i\sin\theta$,
the integration above is divided into two terms, one with the real
part $\cos\theta$ and one with the imaginary part $i\sin\theta$.  The
term with $i\sin\theta$ can be easily obtained as
\begin{equation}
  \begin{split}
&\int_0^{2\pi}i\sin\theta
           (\rho_r \cos\theta)
         [\text{Erf} (\rho_r \cos\theta/a)-1]
         \exp(\rho_r^2\cos^2\theta/a^2)d\theta\\
=&-\int_0^{2\pi}i
           \rho_r \cos\theta
         [\text{Erf} (\rho_r \cos\theta/a)-1]
         \exp(\rho_r^2\cos^2\theta/a^2)d\cos\theta\\
=&-\left(\int_1^{-1}+\int_{-1}^{1}\right)
         i\rho_r t
         [\text{Erf} (\rho_r t/a)-1]
         \exp(\rho_r^2t^2/a^2)dt\\
=&0.
  \end{split}
\end{equation}
Now the remaining term in $f_{n, xy}(r_{n,x}, r_{n,y})$ is
\begin{equation}
  \begin{split}
f_{n, xy}(r_{n,x}, r_{n,y})
\propto \int_0^{2\pi} &
 \exp\left\{-\frac{1}{2 b^2}\rho_r^2 + i\phi_r \right\}
           \frac{1}{\rho_r}
           (\rho_r \cos\theta/a)^2\\
 &\times     
         [\text{Erf} (\rho_r \cos\theta/a)-1]
         \exp(\rho_r^2\cos^2 \theta/a^2)d\theta.\\
  \end{split}
\end{equation}
The integration over $\theta$ can be obtained analytically with Mathematica as
\begin{equation}\label{eq:intTheta}
  \begin{split}
          &\int_0^{2\pi}           
           (\rho_r \cos\theta/a)^2     
         [\text{Erf} (\rho_r \cos\theta/a)-1]
         \exp(\rho_r^2\cos^2 \theta/a^2)d\theta\\
=&-\frac{\pi\rho_r^2}{a^2}\exp\left\{\frac{\rho_r^2}{(2a^2)}\right\}
   \left[I_0\left(\frac{\rho_r^2}{2a^2}\right)+I_1\left(\frac{\rho_r^2}{2a^2}
   \right)\right],
  \end{split}
\end{equation}
where $I_0$ and $I_1$ are the modified Bessel functions of the first
kind. We will show the proof of Eq.~(\ref{eq:intTheta}) later.
Substitute this equation into $f_{n, xy}(r_{n,x}, r_{n,y})$ and we
have
\begin{equation}
  \begin{split}
& f_{n, xy}(r_{n,x}, r_{n,y})
\propto \exp\left\{(\frac{1}{2 a^2}-\frac{1}{2 b^2})\rho_r^2 + i\phi_r \right\}
    \rho_r
   \left[I_0\left(\frac{\rho_r^2}{2a^2}\right)+I_1\left(\frac{\rho_r^2}{2a^2}
   \right)\right].
  \end{split}
\end{equation}
which can be written as
 \begin{equation}
   \begin{split}
f_{n, xy}(r_{n,x}, r_{n,y}) \propto
   &\exp \left\{-\frac{r_{n,x}^2+r_{n,y}^2}{C_{n,xy}^2}
                \right\}\\
   &\times [I_0(\frac{r_{n,x}^2+r_{n,y}^2}{A^2})
            +I_1(\frac{r_{n,x}^2+r_{n,y}^2}{A^2})]\\
   &\times (r_{n,x}^2+r_{n,y}^2)^{1/2} e^{i \phi_{r_n}},
   \end{split}
\end{equation}          
where 
 \begin{equation}
   \begin{split}
C_{n,xy}^2=-\left( \frac{1}{2a^2}-\frac{1}{2b^2} \right)^{-1}
=\frac{8b^4+4b^2\beta_{n,xy}^2}{4b^2+\beta_{n,xy}^2},
\end{split}
\end{equation} 
and
 \begin{equation}
   \begin{split}
A^2=2a^2=4b^2+ \frac{8b^4}{\beta_{n,xy}^2}.
\end{split}
\end{equation} 
Thus we have obtained analytically all the integration results of the
generate coordinates , and the THSR wave function can now be written
in the single function form as,
 \begin{equation}
   \langle \mathbf{r}_1,\sigma_1,\tau_1,\dots,\mathbf{r}_n,\sigma_n,\tau_n
   \left| \Phi \right\rangle \propto \mathcal{A}
      \left[ F(\mathbf{X}_1,\mathbf{X}_2, \mathbf{r}_n)
        \phi(\alpha_1)\phi(\alpha_2) 
      \right]
 \end{equation}
where
 \begin{equation}
   \begin{split}
 F(\mathbf{X}_1,\mathbf{X}_2, \mathbf{r}_n)   
 =&F_\alpha(\mathbf{R}_1, \mathbf{R}_2)f_{n, xy}(r_{n,x}, r_{n,y})f_{n, z}(r_{n,z})\\
  = &\exp \left\{-\frac{X_{1,x}^2+X_{1,y}^2+X_{2,x}^2+X_{2,y}^2}{C_{\alpha,xy}^2}
                      -\frac{X_{1,z}^2+X_{2,z}^2}{C_{\alpha,z}^2}
         \right\}  \\
   &\times \exp \left\{-\frac{r_{n,x}^2+r_{n,y}^2}{C_{n,xy}^2}
                      -\frac{r_{n,z}^2}{C_{n,z}^2}
                \right\}\\
   &\times [I_0(\frac{r_{n,x}^2+r_{n,y}^2}{A^2})
            +I_1(\frac{r_{n,x}^2+r_{n,y}^2}{A^2})]\\
   &\times (r_{n,x}^2+r_{n,y}^2)^{1/2} e^{\pm i \phi_{r_n}}
   \end{split}
 \end{equation}    
 where~ $C_{\alpha,xy}^2 = b^2/2+\beta_{\alpha,xy}^2$,~
 $C_{\alpha,z}^2 = b^2/2+\beta_{\alpha,z}^2$,~ $C_{n,xy}^2 =
 (8b^4+4b^2\beta_{n,xy}^2)/(4b^2+\beta_{n,xy}^2)$,~ $C_{n,z}^2 = 2
 b^2+\beta_{n,z}^2$, ~and~ $A^2 =4b^2+ 8b^4/\beta_{n,xy}^2$.
 
\subsection*{Proof of E\MakeLowercase{q}.~(A.12) in the Appendix}
We will apply the substitution $t=\rho_r/a$ in Eq.~(A.12) and prove
the following integral,
\begin{equation}\label{eq:integral}
  \begin{split}
\int_0^{2\pi} &t^2 \cos^2\theta[ \text{Erf} (t \cos\theta)-1]
    \exp(t^2\cos^2\theta)d\theta\\
&=-\pi t^2\exp\left(\frac{t^2}{2}\right)
    \left[I_0\left(\frac{t^2}{2}\right)
    +I_1\left(\frac{t^2}{2}\right)\right],
  \end{split}
\end{equation}
First we evaluate the integral of the first term in $[\cdots]$ in the
left side in Eq.~(\ref{eq:integral}):
\begin{equation}\label{eq:firstTerm}
  \begin{split}
\int_0^{2\pi} &t^2 \cos^2\theta \text{Erf} (t \cos\theta)
    \exp(t^2\cos^2\theta)d\theta\\
&=\int_0^{\pi/2}\cdots d\theta+\int_{\pi/2}^{\pi}\cdots d\theta
  +\int_\pi^{3\pi/2}\cdots d\theta+\int_{3\pi/2}^{2\pi}\cdots d\theta.
  \end{split}
\end{equation}
The third term in Eq.~(\ref{eq:firstTerm}) can be written as
\begin{equation}
  \begin{split}
\int_{\pi}^{3\pi/2} &t^2 \cos^2\theta\text{Erf} (t \cos\theta)
    \exp(t^2\cos^2\theta)d\theta\\
&=\int_0^{\pi/2}t^2 \cos^2\theta\text{Erf} (-t \cos\theta)
    \exp(t^2\cos^2\theta)d\theta\\
&=-\int_0^{\pi/2}t^2 \cos^2\theta\text{Erf} (t \cos\theta)
    \exp(t^2\cos^2\theta)d\theta.\\
  \end{split}
\end{equation}
Thus the sum of the first and third terms in Eq.~(\ref{eq:firstTerm})
is zero. In the same manner the sum of the second and fourth terms in
Eq.~(\ref{eq:firstTerm}) is zero. Then the result of
Eq.~(\ref{eq:firstTerm}) can be written as
\begin{equation}
\int_0^{2\pi} t^2 \cos^2\theta \text{Erf} (t \cos\theta)
    \exp(t^2\cos^2\theta)d\theta =0.
\end{equation}
Thus we have to prove the following equation:
\begin{equation}\label{eq:secondTerm}
\int_0^{2\pi} \cos^2\theta \exp(t^2\cos^2\theta)d\theta 
=\pi \exp\left(\frac{t^2}{2}\right)
    \left[I_0\left(\frac{t^2}{2}\right)
    +I_1\left(\frac{t^2}{2}\right)\right].
\end{equation}
With use of a simple formula, $\cos^2\theta = (\cos 2\theta + 1)/2$,
Equation (\ref{eq:secondTerm}) can be written as
\begin{equation}\label{eq:with2Theta}
\int_0^{2\pi} (1+\cos 2\theta) \exp \left(\frac{t^2}{2}\cos 2\theta \right)d\theta 
=2\pi \left[I_0\left(\frac{t^2}{2}\right)
    +I_1\left(\frac{t^2}{2}\right)\right].
\end{equation}
Next we will prove Eq.~(\ref{eq:with2Theta}). For simplicity, we apply
the substitution of $z = t^2 /2$.
\begin{equation}\label{eq:withZ}
\begin{split}
\int_0^{2\pi} \exp \left(z \cos 2\theta \right)d\theta &
=\int_0^{2\pi} \exp \left(z \cos \theta \right)d\theta \\
&=\sum_{n=0}^{\infty} \frac{z^n}{n!} \int_0^{2\pi} (\cos \theta)^n d\theta
\end{split}
\end{equation}
With use of the following formula,
\begin{align}
&\int_0^{2\pi} (\cos \theta)^n d\theta
\left\{
\begin{array}{cl}
2\sqrt{\pi}\,\Gamma(k+\frac{1}{2})/\Gamma(k+1) &\text{for } n=2k\\
0                                              &\text{for } n=\text{odd}\\
\end{array}\right.,\\
&\Gamma\left(k+\frac{1}{2}\right)=\frac{\sqrt{\pi}(2k)!}{2^{2k}k!}\\
&I_\nu(z)=\left(\frac{z}{2}\right)^\nu
    \sum_{k=0}^\infty\frac{1}{k!\Gamma(\nu+k+1)}
    \left(\frac{z}{2}\right)^{2k},
\end{align}
we can write down as follows,
\begin{equation}
\begin{split}
\int_0^{2\pi}& \exp(z \cos\theta) d\theta\\
&=\sum_{k=0}^\infty \frac{z^{2k}}{(2k)!}\int_0^{2\pi} (\cos\theta)^{2k} d\theta\\
&=2\pi \sum_{k=0}^\infty\frac{1}{k!\Gamma(k+1)}
    \left(\frac{z}{2}\right)^{2k}\\
&=2\pi I_0(z).
\end{split}
\end{equation}
On the other hand,
\begin{equation}
\begin{split}
\int_0^{2\pi}& \cos 2\theta \exp(z \cos 2\theta) d\theta\\
&=\int_0^{2\pi} \cos \theta \exp(z \cos \theta) d\theta\\
&=\sum_{n=0}^\infty \frac{z^{n}}{n!}\int_0^{2\pi} (\cos\theta)^{n+1} d\theta\\
&=\sum_{k=0}^\infty \frac{z^{2k+1}}{(2k+1)!}\int_0^{2\pi} (\cos\theta)^{2k+2} d\theta\\
&=2\pi\times\left(\frac{z}{2}\right) \sum_{k=0}^\infty\frac{1}{k!\Gamma(k+2)}
    \left(\frac{z}{2}\right)^{2k}\\
&=2\pi I_1(z).
\end{split}
\end{equation}
Then Eqs.~(\ref{eq:with2Theta}) and (\ref{eq:integral}) were proved.

\end{document}